\documentclass[10pt,conference]{IEEEtran}

\usepackage{amsmath,amssymb,amsfonts}
\usepackage{graphicx}
\usepackage{textcomp}
\usepackage[table]{xcolor}
\usepackage{algorithm}
\usepackage{algorithmic}

\begin{document}

\title{Decoder-Prior Poisoning in Quantum Error Correction: Attacks and PriorGuard Defense}

\author{\IEEEauthorblockN{Xinyi Li}
\IEEEauthorblockA{\textit{Stevens Institute of Technology}\\
Hoboken, USA \\
xli215@stevens.edu}
\and
\IEEEauthorblockN{Yifeng Peng}
\IEEEauthorblockA{\textit{Stevens Institute of Technology}\\
Hoboken, USA \\
ypeng21@stevens.edu}
\and
\IEEEauthorblockN{Juntao Chen}
\IEEEauthorblockA{\textit{Fordham University}\\
New York, USA \\
juntao.chen@nyu.edu}
\and
\IEEEauthorblockN{Ying Wang}
\IEEEauthorblockA{\textit{Stevens Institute of Technology}\\
Hoboken, USA \\
ywang6@stevens.edu}
}

\maketitle

\begin{abstract}
Quantum error correction (QEC) protects quantum computations by repeatedly measuring stabilizer syndromes and using a classical decoder to infer corrections. Modern surface-code decoders are increasingly calibration-aware: they use recent device behavior to set priors such as matching-graph edge probabilities, and these priors directly shape the selected correction. We identify the prior-update path as an overlooked integrity-critical attack surface: a poisoned prior can change decoding decisions even when the syndrome stream, logical labels, and decoder implementation are unchanged. This makes prior poisoning different from ordinary syndrome anomalies and difficult for raw syndrome-anomaly tests to detect. We introduce PriorGuard, a lightweight semantic guard that checks proposed prior changes on high-influence detector-graph entries against private syndrome-derived evidence, accepts updates only when their log-odds movement is evidence-supported, and otherwise falls back to the last evidence-supported prior. We evaluate on Stim-generated rotated surface-code memory circuits decoded with PyMatching. Prior influence is highly concentrated, with the top 10\% of entries carrying over 60\% of positive logical-error influence; poisoning only 6.4\% of decoder-prior entries can raise logical error rate (LER) by 3.35$\times$. PriorGuard recovers most attack-induced loss with a low false-positive rate and modest update-boundary overhead of about 1.7--1.8 ms and less than 0.31 MB auxiliary memory per update.
\end{abstract}


\section{Introduction}

Quantum noise can degrade quantum-computing workloads~\cite{peng2026quantum},
while surface-code quantum error correction provides a leading path toward
scalable fault tolerance~\cite{google2025quantum}. Practical decoders increasingly set priors from recent device behavior rather than a fixed noise model~\cite{sivak2024optimization}. In matching-based decoding, these priors specify edge probabilities that are converted into weights used to determine the correction.

\begin{figure}[t]
  \centering
  \includegraphics[width=0.95\linewidth]{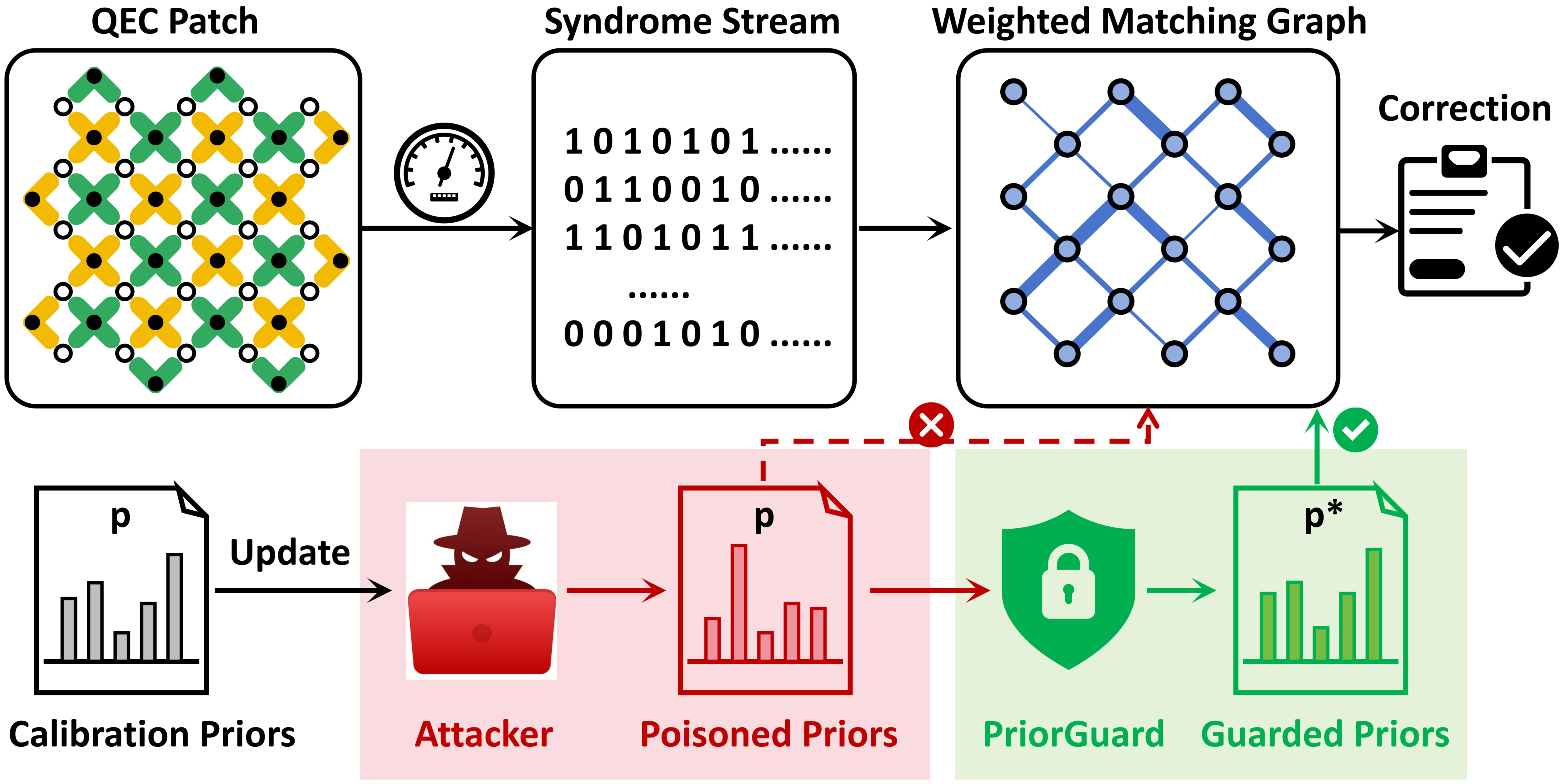}
  \caption{Threat model for decoder-prior poisoning. The syndrome path is trusted, while the prior-update path is untrusted and guarded before decoding.}
  \label{fig:threat-model-pipeline}
\end{figure}

This prior-update path is rarely treated as a security boundary. Published quantum-security work has focused on malicious-circuit screening and hardware-facing defenses~\cite{deshpande2023design}, while recent calibration-aware QEC work treats priors as accuracy-tuning parameters~\cite{sivak2024optimization}. Calibration-to-decoder priors therefore remain unguarded. As shown in Fig.~\ref{fig:threat-model-pipeline}, measured syndromes follow the trusted decoding path, while calibration-derived priors enter the weighted matching graph through a separate update channel. A poisoned calibration update can therefore redirect decoding while the syndrome-extraction circuit, logical labels, and decoder implementation remain unchanged. This makes prior poisoning a concrete integrity problem for calibration-aware decoding. By altering priors later consumed by the matching decoder, the adversary can compromise the protected computation and increase the chance of an incorrect logical outcome.

\begin{figure}[t]
  \centering
  \includegraphics[width=0.95\linewidth]{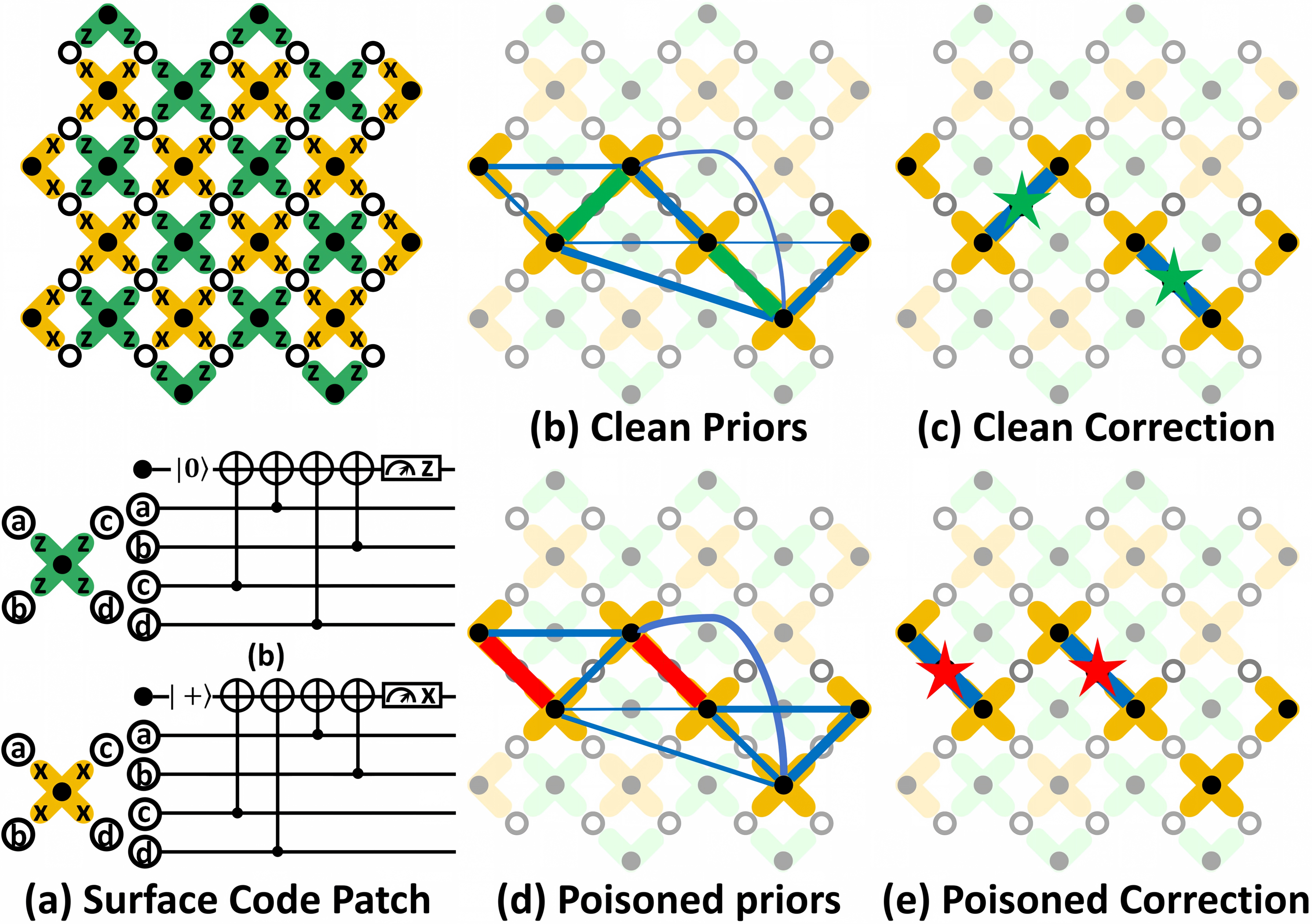}
  \caption{Prior poisoning mechanism. (a) Surface-code patch, (b) clean priors, (c) clean correction, (d) poisoned priors, and (e) poisoned correction.}
  \label{fig:prior-poisoning-correction}
\end{figure}

The syndrome identifies detector events, but the prior determines how the matching decoder scores candidate corrections. A corrupted prior can therefore change the selected correction even for the same syndrome. The mechanism is shown in Fig.~\ref{fig:prior-poisoning-correction}. Clean priors yield one correction in Fig.~\ref{fig:prior-poisoning-correction}(b),(c), while poisoned priors can make the same syndrome favor a different weighted path and correction in Fig.~\ref{fig:prior-poisoning-correction}(d),(e). The changed correction may differ from the clean correction by a logical operator. Thus, a prior update directly affects the logical decision made by the decoder~\cite{darmawan2025optimal}.

Existing research guards other parts of this stack while leaving the prior-update channel largely unchecked. Online noise estimation and recalibration re-derive priors from syndrome statistics to track drift~\cite{kobori2025bayesian}, and reliability-oriented decoding gates accuracy by re-running decoders~\cite{shutty2026efficient}. Both reason over syndromes and outputs computed from the same priors, so a prior-only attack that keeps syndromes plausible passes through. Prior optimization treats prior quality as an accuracy objective~\cite{sivak2024optimization,darmawan2025optimal}, and quantum-security defenses screen circuits or harden the execution stack~\cite{deshpande2023design}. The missing primitive is semantic prior-integrity checking, which accepts a prior update only when trusted evidence supports the change.

PriorGuard implements this prior-integrity check as an update-boundary guard. Given a proposed prior update, it monitors influence-ranked decoder-graph entries, estimates recent syndrome-derived evidence for those entries, and compares the proposed log-odds movement against a normal-update threshold. Updates inconsistent with the evidence are rejected before they reach the matching decoder. Our contributions are summarized as follows:
\begin{itemize}
  \item To our knowledge, we are the first to frame the calibration-derived prior-update channel as an integrity-critical attack surface in quantum decoding.
  \item We demonstrate prior-only poisoning attacks that leave syndromes unchanged yet substantially raise the logical error rate (LER).
  \item We introduce PriorGuard, an influence-guided evidence-consistency layer for matching-graph prior updates.
  \item We provide circuit-derived evidence that poisoning only \(6.4\%\) of decoder-prior entries can raise LER by \(3.35\times\), while PriorGuard recovers most attack-induced loss with about \(1.7\) to \(1.8\) ms/update latency and less than \(0.31\) MB auxiliary memory.
\end{itemize}

\section{Threat Model}

We consider an adversary in the classical calibration-to-decoder prior path. The adversary sits after calibration artifacts are produced and before the matching decoder consumes a prior update, as shown in Fig.~\ref{fig:threat-model-pipeline}. Realistic roles include a compromised calibration job, outsourced decoder or control service, untrusted storage or logging component, co-tenant with artifact write access, or malicious insider. This position is plausible in deployments that separate quantum measurement, calibration, and decoding services.

The adversary can observe detector-graph edge identifiers, the prior schema, previously released priors, and update accept or reject outcomes. It can submit, modify, or replay prior values for selected matching edges. The trust boundary contains the syndrome-extraction circuit, raw measurement-to-syndrome path, observable labels, decoder implementation, and private audit syndrome windows. The adversary cannot change syndromes, labels, the circuit, or decoder code, and cannot read the guard's held-out audit samples.

An attack succeeds when an unsupported prior update changes the decoder's correction behavior or measurably raises LER while all trusted decoding inputs remain valid. This model focuses on the calibration-to-decoder interface because this is where a classical artifact is promoted into decoder weights and can alter logical outcomes without touching the measurement path.

\section{PriorGuard Model}

\begin{figure}[!t]
  \centering
  \includegraphics[width=0.95\linewidth]{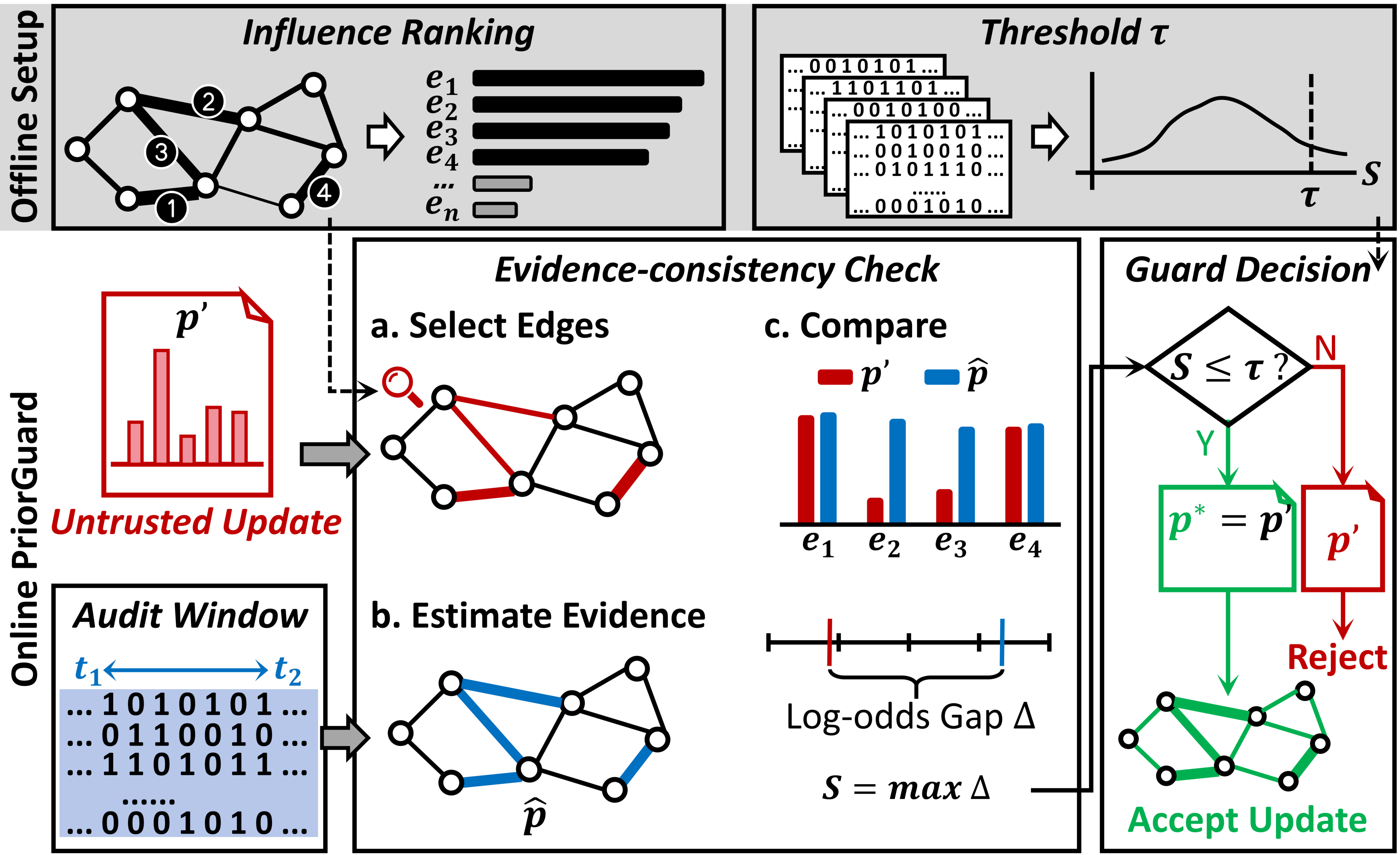}
  \caption{PriorGuard workflow for evidence-consistent prior updates.}
  \label{fig:priorguard-method}
\end{figure}

PriorGuard protects the prior-update boundary with three modules, as shown in Fig.~\ref{fig:priorguard-method}: offline profiling, an evidence-consistency check, and a guard decision. Offline profiling prepares fixed monitoring metadata. The online gate checks each proposed prior update against private audit evidence and emits a guarded prior vector for decoder consumption. The following paragraphs describe these modules in order.

\begin{algorithm}[t]
\caption{PriorGuard Prior-Update Gate}
\label{alg:priorguard}
\begin{algorithmic}[1]
\REQUIRE proposed prior \(p'\); audit window of \(N\) shots; monitor set \(M\); threshold \(\tau\); fallback prior \(p_{\mathrm{fb}}\)
\ENSURE guarded prior \(p^{*}\) consumed by the decoder
\FORALL{monitored edges \(i\in M\)}
  \STATE \(c_i \leftarrow\) endpoint co-firings of edge \(i\) in the audit window
  \STATE \(\hat p_i \leftarrow (c_i+\alpha)/(N+2\alpha)\)
  \STATE \(\delta_i \leftarrow |\ell(p'_i)-\ell(\hat p_i)|\)
\ENDFOR
\STATE \(S \leftarrow \max_{i\in M}\delta_i\) \COMMENT{monitored consistency score}
\IF{\(S \le \tau\)}
  \STATE \(p^{*} \leftarrow p'\) \COMMENT{evidence-supported: accept update}
\ELSE
  \STATE \(p^{*} \leftarrow p_{\mathrm{fb}}\) \COMMENT{unsupported: reject, keep last-good prior}
\ENDIF
\STATE \textbf{return} \(p^{*}\)
\end{algorithmic}
\end{algorithm}

\paragraph{Offline profiling}
For each decoder-graph class, PriorGuard ranks candidate priors by logical influence on a held-out corpus disjoint from audit and test data. It perturbs one edge prior at a time, re-decodes the same corpus, and ranks entries by the induced \(\Delta\)LER. This step identifies which prior entries are security-critical under a fixed monitoring budget, influenced by~\cite{peng2026titan}. The monitor set \(M\) is the high-influence head of this ranking. PriorGuard also calibrates a threshold \(\tau\) from normal update windows, so ordinary calibration noise and expected drift define the accepted score range.

\paragraph{Evidence-consistency check}
When a proposed update \(p'\) arrives, PriorGuard draws a private audit window of \(N\) shots from the trusted syndrome stream and checks only monitored entries. For an interior edge \(i=(u,v)\), \(c_i\) counts joint firings of detectors \(u\) and \(v\); for a boundary edge, it counts firing of the finite endpoint. With smoothing parameter \(\alpha\),
\[
  \hat p_i=\frac{c_i+\alpha}{N+2\alpha}.
\]
This evidence is trusted but noisy; it is used to validate \(p'\), not to replace the full prior vector or re-run calibration.
The check compares proposed and evidence-derived priors in log-odds space,
\[
  \ell(p)=\log_{10}\frac{p}{1-p},
\]
because matching weights are monotone in probability odds. With \(\delta_i=|\ell(p'_i)-\ell(\hat p_i)|\), the consistency score is
\[
  S=\max_{i\in M} \delta_i .
\]

\paragraph{Guard decision}
The guard decision consumes the consistency score \(S\) and the normal-update threshold \(\tau\). If \(S\le\tau\), the update is accepted and \(p^*=p'\). Otherwise, the update is unsupported by audit evidence and PriorGuard emits a fallback guarded prior \(p^*=p_{\mathrm{fb}}\), such as the last evidence-supported prior. The emitted \(p^*\) becomes the decoder-consumable prior vector for that update.

\begin{figure}[t]
  \centering
  \includegraphics[width=\linewidth]{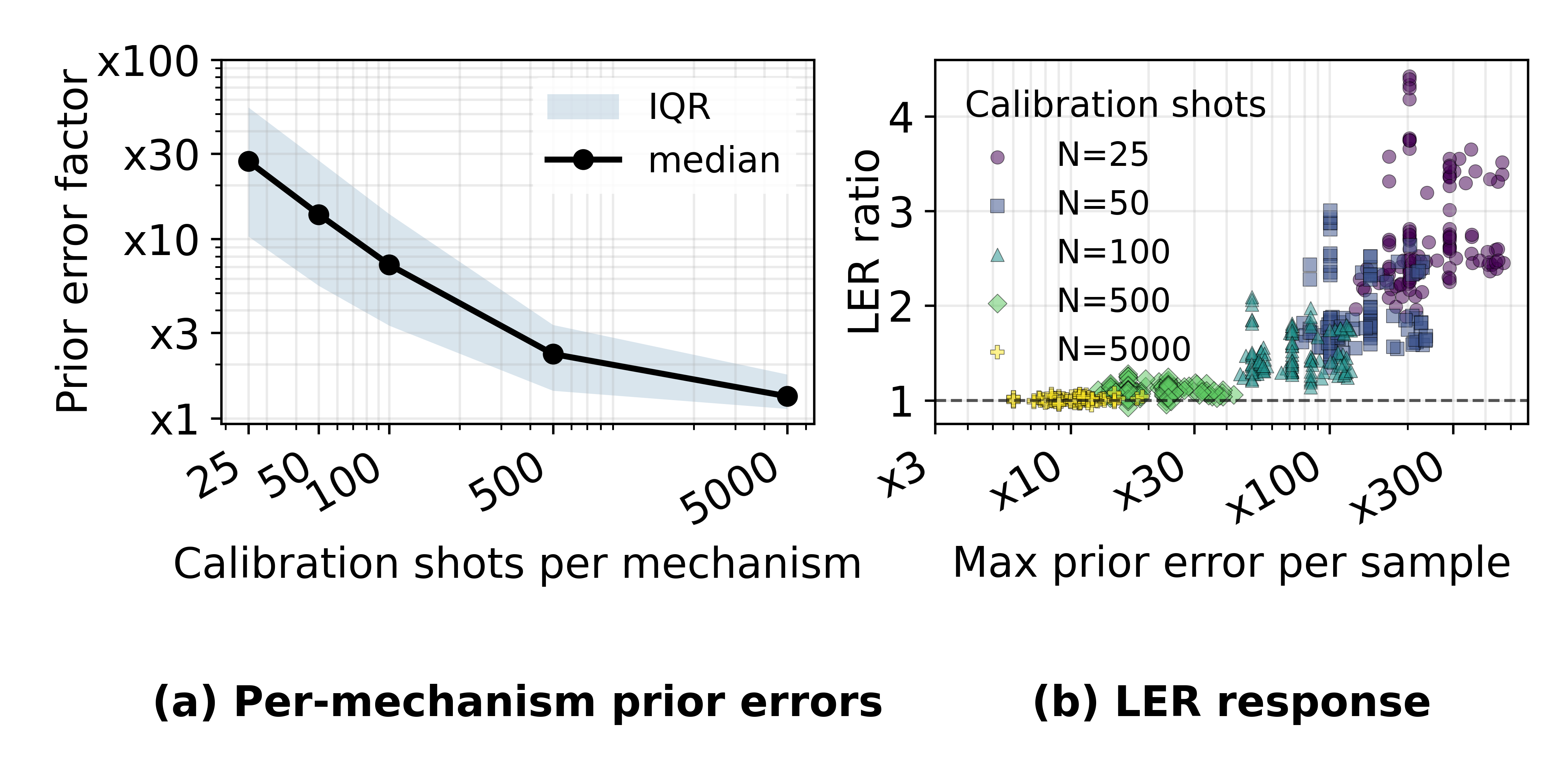}
  \caption{Natural prior uncertainty under a calibration-shot sweep \(N\in\{25,50,100,500,5000\}\): (a) per-mechanism finite-shot prior-error factor; (b) LER ratio versus per-sample maximum prior error.}
  \label{fig:e1-natural-prior-uncertainty}
\end{figure}

\section{Evaluation and Results}
\label{sec:eval}
\paragraph{Experimental Setup}
\textbf{Benchmarks.} We evaluate on Stim-generated rotated surface-code memory circuits with \(d\in\{5,7,9\}\), \(d\) syndrome rounds, and base circuit-level noise probability \(0.0045\). Each circuit is decoded with PyMatching from a decomposed detector error model (DEM), and decoder priors are evaluated under heterogeneous, biased, and slow-drift noise transforms. \textbf{Baselines.} We compare against signed-only authentication, random monitoring, and global clipping.

\paragraph{Natural prior fluctuations}
Routine calibration naturally produces statistical variation in decoder priors, so prior movement alone does not indicate security-relevant tampering. Fig.~\ref{fig:e1-natural-prior-uncertainty}(a) simulates finite-shot calibration by sampling noisy prior estimates around the underlying DEM rates, and Fig.~\ref{fig:e1-natural-prior-uncertainty}(b) decodes the same fixed syndrome corpora with those estimates. Fewer calibration shots produce larger odds error and can increase LER, showing that priors are noisy estimates rather than fixed truth values. A guard must therefore tolerate ordinary fluctuations while rejecting unsupported changes.

\begin{figure}[t]
  \centering
  \includegraphics[width=\linewidth]{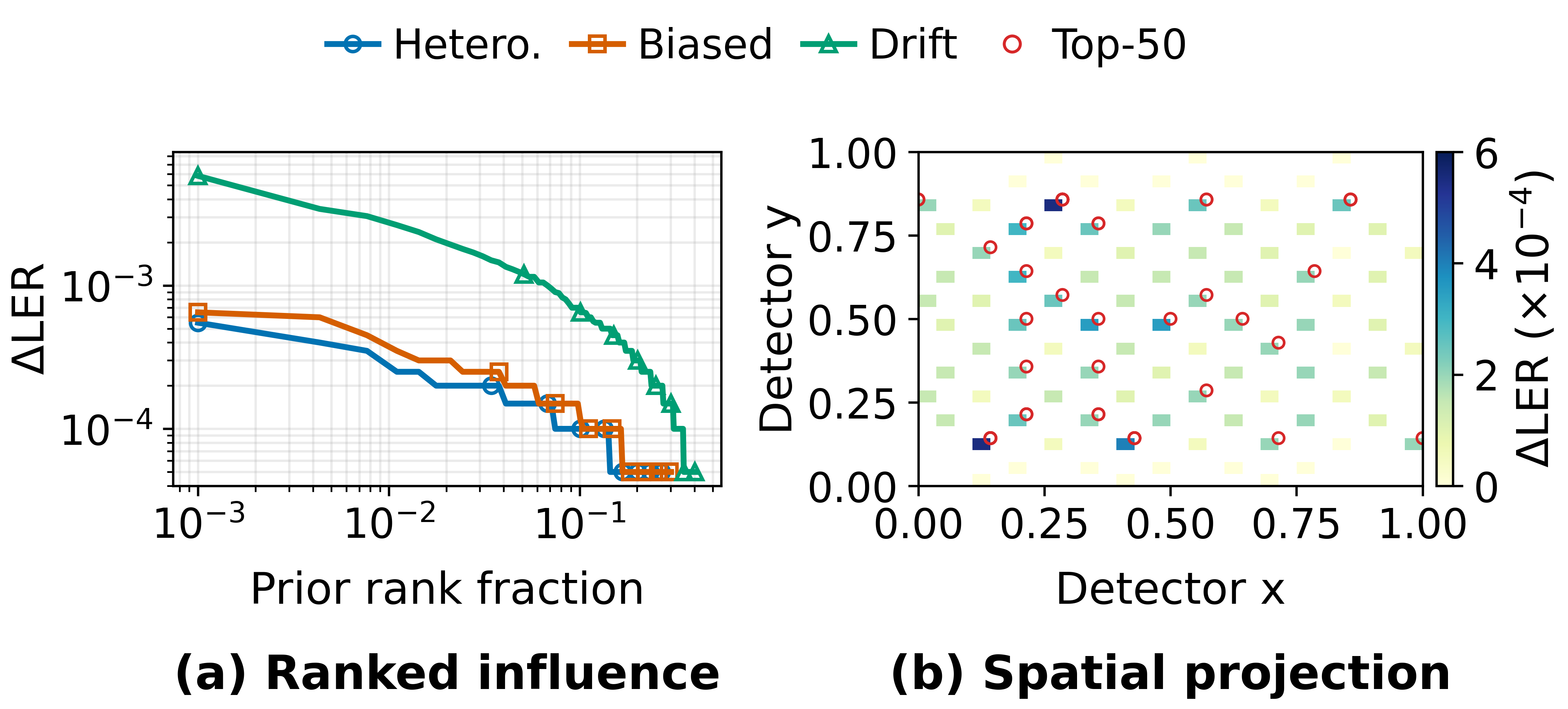}
  \caption{Prior-influence landscape: (a) ranked single-prior LER increase \(\Delta\)LER; (b) detector-graph locations of high-\(\Delta\)LER priors. White bins have no projected candidate edge.}
  \label{fig:e2a-prior-influence-landscape}
\end{figure}

\paragraph{Prior influence is concentrated}
We next ask whether all prior entries are equally security-critical. Single-edge ablation ranks candidate PyMatching edges by the induced \(\Delta\)LER after multiplying one prior odds value and re-decoding the same corpus. Fig.~\ref{fig:e2a-prior-influence-landscape}(a) shows a steep heavy tail, and Fig.~\ref{fig:e2a-prior-influence-landscape}(b) localizes high-impact entries in a sparse projected detector-graph subset. Table~\ref{tab:e2a-influence-concentration} shows that the top \(10\%\) of entries capture more than \(60\%\) of positive \(\Delta\)LER, motivating influence-guided monitoring for decoder-prior integrity.

\begin{table}[t]
  \centering
  \caption{Concentration of positive \(\Delta\)LER in top-ranked prior edges.}
  \label{tab:e2a-influence-concentration}
  \footnotesize
  \setlength{\tabcolsep}{0pt}
  \renewcommand{\arraystretch}{1.08}
  \begin{tabular}{@{}p{0.34\linewidth}p{0.22\linewidth}p{0.22\linewidth}p{0.22\linewidth}@{}}
    \hline
    \rowcolor[gray]{0.90}
    Noise model & Top 1\% & Top 5\% & Top 10\% \\
    \hline
    Heterogeneous & 13.8\% & 40.5\% & 61.6\% \\
    Biased & 14.4\% & 41.7\% & 63.4\% \\
    Slow drift & 16.2\% & 46.4\% & 66.7\% \\
    \hline
  \end{tabular}
\end{table}

\paragraph{Prior-only poisoning is damaging}
The matched-budget poisoning study tests whether this concentration is exploitable while syndromes, labels, and decoder code remain fixed. Table~\ref{tab:e2b-matched-budget-poisoning} shows that an odds-\(10\times\) attack on \(k=100\) high-influence entries, only \(6.4\%\) of priors, raises LER by \(3.35\times\), compared with \(1.38\times\) for random poisoning. Thus prior poisoning is not merely calibration noise: targeted corruption of a small prior subset can create disproportionate logical damage.

\begin{table}[t]
  \caption{Matched-budget prior poisoning. Ratios are relative to clean LER; gain is high-influence over random control.}
  \label{tab:e2b-matched-budget-poisoning}
  \centering
  \scriptsize
  \setlength{\tabcolsep}{0pt}
  \renewcommand{\arraystretch}{1.08}
  \begin{tabular*}{\linewidth}{@{\extracolsep{\fill}}ccccc@{}}
    \hline
    \rowcolor[gray]{0.90}
    \multicolumn{5}{c}{Budget sweep, odds-\(10\times\)} \\
    \hline
    Poisoned entries \(k\) & Affected priors & Random & High-inf. & Gain \\
    \hline
    10 & 0.6\% & \(1.04\times\) & \(1.62\times\) & \(\boldsymbol{1.56\times}\) \\
    50 & 3.2\% & \(1.14\times\) & \(2.87\times\) & \(\boldsymbol{2.53\times}\) \\
    100 & 6.4\% & \(1.38\times\) & \(3.35\times\) & \(\boldsymbol{2.42\times}\) \\
    500 & 32.1\% & \(2.96\times\) & \(3.09\times\) & \(\boldsymbol{1.04\times}\) \\
    1000 & 64.2\% & \(3.61\times\) & \(2.71\times\) & \(0.75\times\) \\
    1558 & 100\% & \(2.26\times\) & \(2.26\times\) & \(1.00\times\) \\
    \hline
    \rowcolor[gray]{0.90}
    \multicolumn{5}{c}{Magnitude sweep, \(k=100\)} \\
    \hline
    Prior odds mult. & Affected priors & Random & High-inf. & Gain \\
    \hline
    \(2\times\) & 6.4\% & \(1.02\times\) & \(1.14\times\) & \(\boldsymbol{1.12\times}\) \\
    \(3\times\) & 6.4\% & \(1.06\times\) & \(1.38\times\) & \(\boldsymbol{1.31\times}\) \\
    \(10\times\) & 6.4\% & \(1.38\times\) & \(3.35\times\) & \(\boldsymbol{2.42\times}\) \\
    \hline
  \end{tabular*}
\end{table}

\begin{figure}[t]
  \centering
  \includegraphics[width=\linewidth]{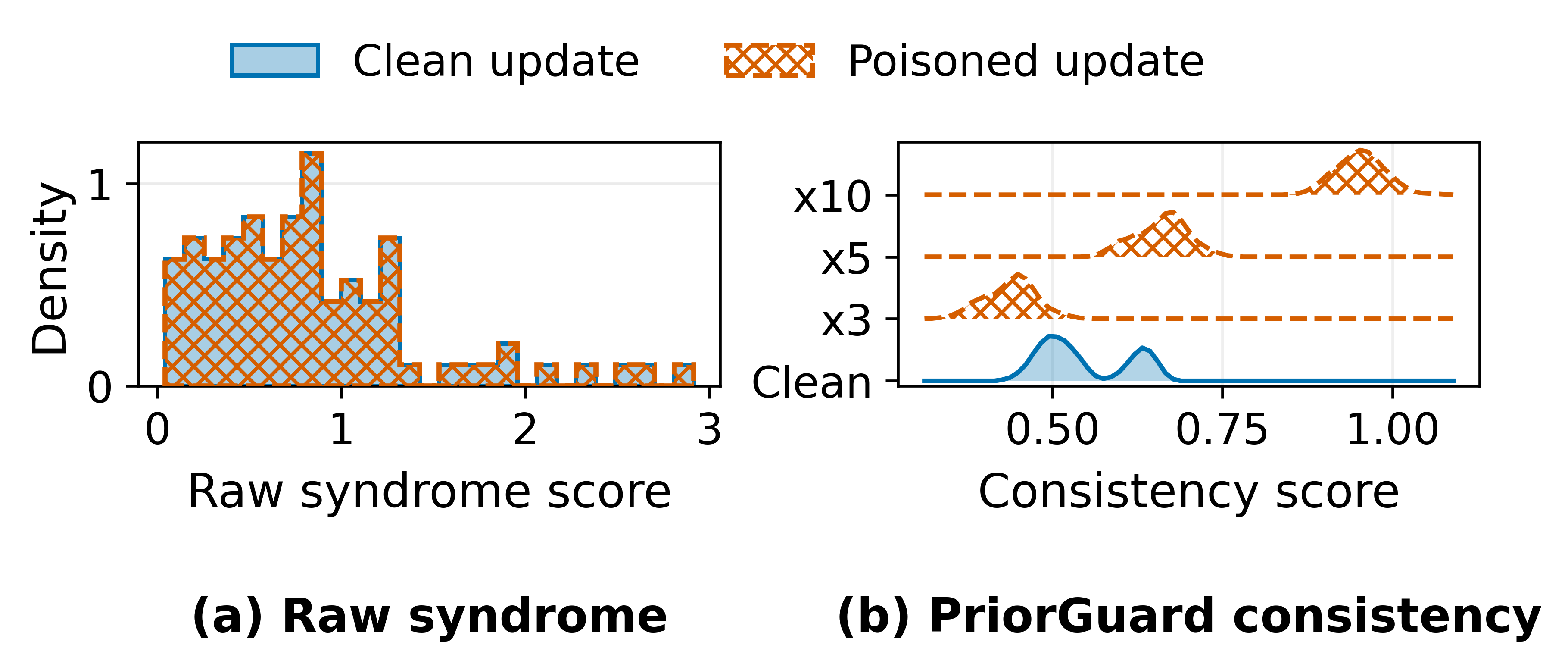}
  \caption{Syndrome plausibility versus prior consistency: (a) raw syndrome anomaly score; (b) PriorGuard consistency score under prior-odds multipliers.}
  \label{fig:e3-syndrome-prior-consistency}
\end{figure}

\paragraph{Syndrome plausibility is insufficient}
Quantum-enhanced anomaly scoring has been studied~\cite{peng2025qsco},
but prior poisoning may leave syndrome statistics unchanged. Thus we then test whether the attack is visible to syndrome-only monitoring. Clean and poisoned prior updates are scored on the same audit syndrome windows, with poisoned priors scaled by \(3\times\), \(5\times\), or \(10\times\). Fig.~\ref{fig:e3-syndrome-prior-consistency}(a) shows overlapping raw syndrome anomaly scores, while Fig.~\ref{fig:e3-syndrome-prior-consistency}(b) shows PriorGuard's consistency score increasing with unsupported perturbation. Prior poisoning is therefore an evidence-consistency violation, not necessarily a syndrome anomaly.

\begin{figure}[t]
  \centering
  \includegraphics[width=0.95\linewidth]{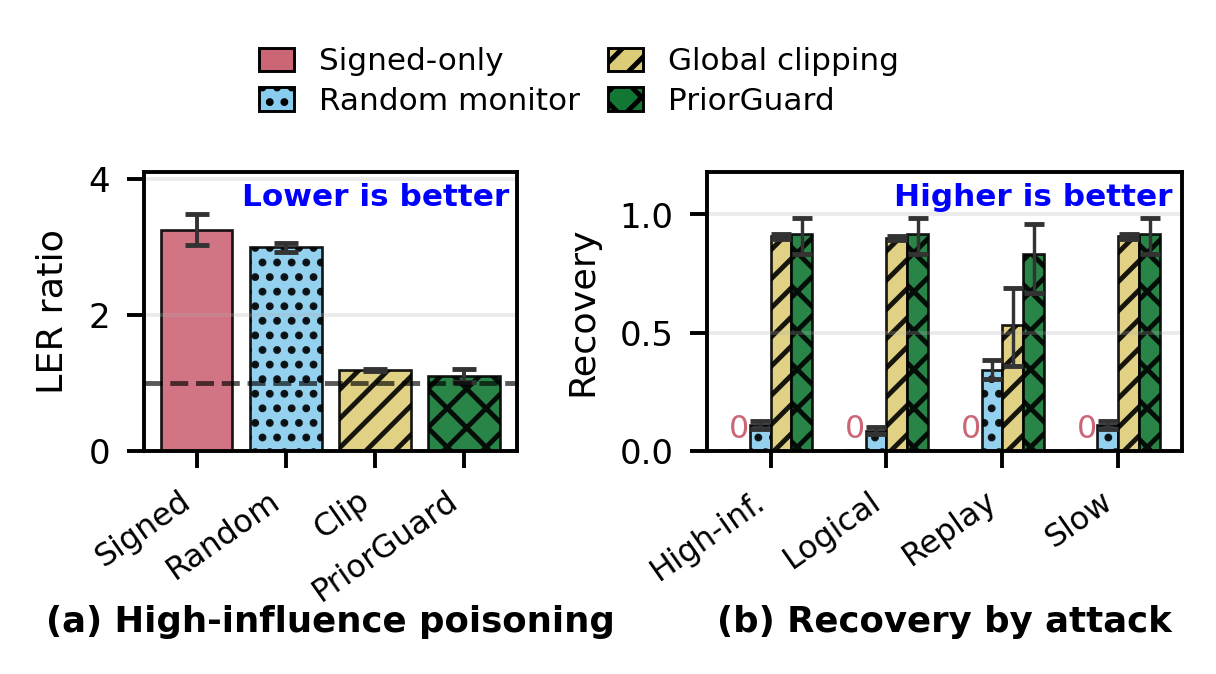}
  \caption{PriorGuard defense efficacy: (a) LER ratio under high-influence poisoning; (b) recovery ratio across attack classes.}
  \label{fig:e4-main-defense}
\end{figure}

\paragraph{PriorGuard recovers LER under attack}
Fig.~\ref{fig:e4-main-defense}(a) compares PriorGuard with baselines under high-influence poisoning, and Fig.~\ref{fig:e4-main-defense}(b) evaluates recovery across separate attack windows. Results show that signed-only authentication accepts poisoned priors, random monitoring can miss high-influence entries, and global clipping raises false positives. PriorGuard restores near-clean LER under high-influence attacks and maintains recovery across logical-path, stale-replay, and slow-poisoning variants via evidence checks on ranked entries.

\begin{figure}[t]
  \centering  
  \includegraphics[width=0.95\linewidth]{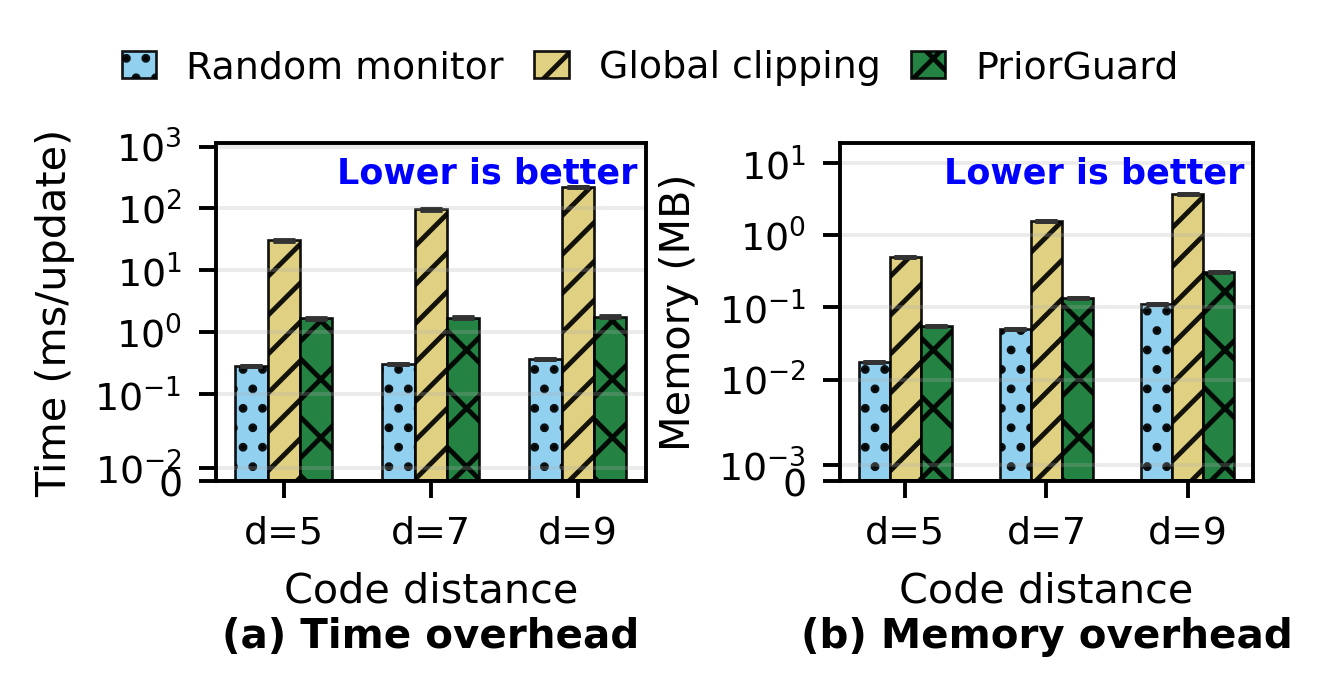}
  \caption{PriorGuard defense overhead: (a) guard-decision latency; (b) peak additional memory.}
  \label{fig:e5-defense-overhead}
\end{figure}

\paragraph{Update-boundary overhead is modest}
Finally, we measure update-boundary cost separately from the final PyMatching decode shared by all methods. Fig.~\ref{fig:e5-defense-overhead} shows that PriorGuard requires about \(1.7\)--\(1.8\) ms per update and less than \(0.31\) MB auxiliary memory across \(d=5,7,9\).

\section{Conclusion}

This paper identifies decoder-prior updates as an integrity-critical attack surface
in calibration-aware QEC. Prior-only poisoning can substantially raise
logical error rates without altering the syndrome stream, while concentrated
prior influence enables targeted attacks and monitoring. PriorGuard validates
proposed changes against recent syndrome evidence on high-influence entries
and falls back on unsupported updates, achieving strong recovery with modest
overhead. Adaptive diffuse poisoning remains a limitation, motivating guards
that combine influence-guided and aggregate consistency checks.

\bibliographystyle{IEEEtran}
\bibliography{reference}

\end{document}